\documentclass[twocolumn,showpacs,preprintnumbers,amsmath,amssymb]{revtex4}

\usepackage{graphicx}
\usepackage{dcolumn}
\usepackage{bm}

\begin{document}

\preprint{IUHET-496}

\title{An Interacting Dark Energy Model for the Expansion History 
of the Universe}

\author{Micheal S. Berger}
 \email{berger@indiana.edu}
\author{Hamed Shojaei}%
 \email{seshojae@indiana.edu}
\affiliation{%
Physics Department, Indiana University, Bloomington, IN 47405, USA}

\date{\today}

\begin{abstract}
We explore a model of interacting dark energy where the dark energy density is
related by the holographic principle
to the Hubble parameter, and the decay of the dark energy into matter
occurs at a rate comparable to the current value of the Hubble parameter.
We find this gives a good fit to the observational data supporting an
accelerating Universe, and the model represents a possible 
alternative interpretation of the expansion history of the Universe.
\end{abstract}

\maketitle

\def\al{\alpha}
\def\be{\beta}
\def\ga{\gamma}
\def\de{\delta}
\def\ep{\epsilon}
\def\ve{\varepsilon}
\def\ze{\zeta}
\def\et{\eta}
\def\th{\theta}
\def\vt{\vartheta}
\def\io{\iota}
\def\ka{\kappa}
\def\la{\lambda}
\def\vpi{\varpi}
\def\rh{\rho}
\def\vr{\varrho}
\def\si{\sigma}
\def\vs{\varsigma}
\def\ta{\tau}
\def\up{\upsilon}
\def\ph{\phi}
\def\vp{\varphi}
\def\ch{\chi}
\def\ps{\psi}
\def\om{\omega}
\def\Ga{\Gamma}
\def\De{\Delta}
\def\Th{\Theta}
\def\La{\Lambda}
\def\Si{\Sigma}
\def\Up{\Upsilon}
\def\Ph{\Phi}
\def\Ps{\Psi}
\def\Om{\Omega}
\def\mn{{\mu\nu}}
\def\cD{{\cal D}}
\def\cF{{\cal F}}
\def\cL{{\cal L}}
\def\cS{{\cal S}}
\def\fr#1#2{{{#1} \over {#2}}}
\def\frac#1#2{\textstyle{{{#1} \over {#2}}}}
\def\pt#1{\phantom{#1}}
\def\prt{\partial}
\def\vev#1{\langle {#1}\rangle}
\def\ket#1{|{#1}\rangle}
\def\bra#1{\langle{#1}|}
\def\amp#1#2{\langle {#1}|{#2} \rangle}
\def\half{{\textstyle{1\over 2}}}
\def\lsim{\mathrel{\rlap{\lower4pt\hbox{\hskip1pt$\sim$}}
    \raise1pt\hbox{$<$}}}
\def\gsim{\mathrel{\rlap{\lower4pt\hbox{\hskip1pt$\sim$}}
    \raise1pt\hbox{$>$}}}
\def\ol#1{\overline{#1}}
\def\Re{\hbox{Re}\,}
\def\Im{\hbox{Im}\,}
\def\etal {{\it et al.}}
\def\slash#1{\not\hbox{\hskip -2pt}{#1}}

\section{Introduction}

Observations of type Ia
supernova indicate that the Universe is making a transition
from a decelerating phase to an accelerating one. The conventional explanation
for this behavior is that a dark energy component is coming to dominate 
over and the matter-dominated phase is giving way to a phase dominated by a 
dark energy component. In fact
a good fit is obtained from the observational data by assuming the cosmological
model ($\Lambda$CDM) involving a cosmological constant $\Lambda$ and cold dark
matter in about the ratios $\Omega_{m,0}=0.3$ and $\Omega_{\Lambda,0}=0.7$.

In this paper we show that one can describe the cosmological data with a
model of dark energy that includes an interaction that effects a transition 
(decay) of dark energy into matter. The model incorporates a holographic 
principle\cite{'tHooft:1993gx,Susskind:1994vu} to determine the
dark energy density of the Universe. The principle relates the dark energy 
scale to the Hubble horizon which has been the subject of speculation for 
applying holographic ideas to 
cosmology\cite{Easther:1999gk,Veneziano:1999ts,Kaloper:1999tt,Brustein:1999ua,Bak:1999hd,Cohen:1998zx,Thomas:2002pq,Bousso:2002ju}. 
This is a common choice for imposing 
holography on cosmology, and it has the most natural thermodynamic 
interpretation. However, this conditon imposed on the dark energy yields an
equation of state that is matter-like\cite{Hsu:2004ri} and therefore 
inconsistent with observational data. Subsequently efforts have been made to 
impose a holographic constraint based on some other physical horizon such
as the particle horizon\cite{Fischler:1998st,Bousso:1999xy} 
or the future event horizon\cite{Li:2004rb,Huang:2004wt,Zhang:2005yz,Myung:2005sv,Myung:2005pw,Zhang:2005hs,Kim:2005at,Li:2006ci} 
or some other physical 
condition\cite{Horava:2000tb,Danielsson:2004xw,Horvat:2004vn,Guberina:2005mp,Elizalde:2005ju,Guberina:2006fy,Ke:2004nw}.

A suitable evolution of the Universe is
obtained when, in addition to the holographic dark energy, 
an interaction (decay of dark energy to matter) is 
assumed, and the decay rate should be set roughly equal to the present value
of the Hubble parameter for a good fit to the expansion history of the 
Universe as determined by the supernova and cosmic microwave
background (CMB) data.

Interacting dark energy has been studied 
previously\cite{Amendola:1999qq,Amendola:2000uh,Zimdahl:2000zm,Balakin:2003tk,Wang:2004cp,Cai:2004dk,Pavon:2005yx,Zimdahl:2005bk,Wang:2005jx,Wang:2005ph,Pavon:2005kr,Berger:2006db,Hu:2006ar}
primarily with a goal of 
explaining the cosmic coincidence problem. A survey of the possible dynamics 
of dark energy can be found in Ref.~\cite{Copeland:2006wr}.
In the presence of an interaction
the dark energy can achieve a stable equilibrium that differs from the usual 
de Sitter case or the approach to the stable equilibrium can be made more 
gradual.
Such models offer the hope of solving the coincidence problem that exists in 
the $\Lambda$CDM model where there is no obvious reason why 
the transition from matter domination to domination by the 
dark energy is occuring during the current epoch. 

This paper presents a simple model that gives an acceptable expansion history
of the Universe in terms of a holographic condition on the dark energy that 
relates its size to the Hubble scale. This is a common choice for imposing 
holography on cosmology, and it has the most natural thermodynamic 
interpretation.
The effective equations of state
of matter and dark energy coincide and behave like cold dark matter (CDM) at 
early times. The transition to behavior like a cosmological constant is 
effected by simply assuming there is a constant decay of dark energy into 
matter. The coincidence problem 
appears in the interacting dark energy model as the choice of 
the scale for the dark energy interaction which must be close to the present 
value of the Hubble parameter.

\section{A Framework for Interacting Dark Energy}
The continuity equations for dark energy and matter are
\begin{eqnarray}
&&\dot{\rho}_\Lambda + 3H(1+w_\Lambda)\rho_\Lambda = -Q\;, \nonumber \\
&&\dot{\rho}_m + 3H(1+w_m)\rho_m = Q\; .
\label{twocomp}
\end{eqnarray}
The interaction is given by the quantity $Q$, and the energy-momentum tensor
remains conserved as long as the same factor $Q$ appears on the right hand
side of each equation. To appeal to our intuition and to facilitate comparison
with the more familiar case of $\Lambda$CDM model, it is useful to define 
effective equations of state that play the role of the native equations of 
state when an interaction is present. Following Ref.~\cite{Kim:2005at},
if we define 
\begin{eqnarray}
w_\Lambda ^{\rm eff}=w_\Lambda+{{\Gamma}\over {3H}}\;, \qquad
w_m ^{\rm eff}=w_m-{1\over r}{{\Gamma}\over {3H}}\;.
\end{eqnarray}
where $\Gamma=Q/\rho_\Lambda$ is a rate, 
the continuity equations can be written in their standard form
\begin{eqnarray}
&&\dot{\rho}_\Lambda + 3H(1+w_\Lambda^{\rm eff})\rho_\Lambda = 0\;, 
\nonumber \\
&&\dot{\rho}_m + 3H(1+w_m^{\rm eff})\rho_m = 0\;
\label{definew}
\end{eqnarray} 
Define as usual
\begin{eqnarray}
\Omega_\Lambda={{8\pi\rho_\Lambda}\over {3M_p^2H^2}}\;, \qquad
\Omega_m={{8\pi\rho_m}\over {3M_p^2H^2}}\;. 
\end{eqnarray}
If we restrict our attention to the flat case,
\begin{eqnarray}
\Omega_\Lambda +\Omega_m=1\;,
\label{flatness}
\end{eqnarray}
one obtains a differential equation
\begin{eqnarray}
{{d\Omega_\Lambda}\over {dx}}=-3\Omega_\Lambda (1-\Omega_\Lambda)
\left [w_\Lambda^{\rm eff}-w_m^{\rm eff}\right ]\;.
\label{diffgen}
\end{eqnarray}
where $x=\ln (a/a_0)$ and $a$ is the Friedmann-Robertson-Walker (FRW) scale
factor, and the ''0'' subscript here and on other quantities indicates the 
parameter value at the present time.
For the $\Lambda$CDM model, where $w_\Lambda^{\rm eff}=w_\Lambda=-1$ and
$w_m^{\rm eff}=w_m=0$, the solution is the familiar one which can drive a
small vacuum energy solution  from a fixed point at $\Omega _\Lambda=0$ 
toward a de Sitter Universe fixed point at
$\Omega _\Lambda=1$. In the presence of a holographic principle and/or an 
interaction, the effective equations of state can depend on 
$x$, and there is the possibility that another fixed point can develop at 
a value of $\Omega_\Lambda$ other than $0$ or $1$.

As shown in Ref.~\cite{Berger:2006db} two physical assumptions are required
to determine the evolution of $\Omega _\Lambda$ and $\Omega_m$. 
These can be chosen from the following list: (a) a ``holographic principle'' 
which specifies $\rho_\Lambda$ in terms of some length scale associated with 
a horizon, (b) an assumption for the 
interaction $\Gamma$, or (c) an assumption about 
the native equations of state
$w_\Lambda$ and $w_m$. The assumptions commonly used for (a) are assumptions 
for the length scale $L_\Lambda$ in 
\begin{eqnarray}
\rho_\Lambda={{3c^2M_p^2}\over {8\pi L_\Lambda^2}}\;.
\label{definel}
\end{eqnarray}
Here the constant $c$ represents an order one constant. Some choices have
been to identify the 
length scale with a physical length such as the Hubble horizon, the particle
horizon, or the future event horizon, or perhaps some other parameter.
The interaction rate in (b) can be specified by an assumption of how 
the ratio of rates $\Gamma /H$ varies as a function of $\Omega _\Lambda$.
The assumption for the equations of state in (c) have been largely confined 
to constant ones.

It is important to note that specifying two of the three conditions determines
the third. For example a generic choice for the length scale $L_\Lambda$ and 
for the interaction rate $\Gamma$ is inconsistent with a constant equation of 
state.
It is easy to show that\cite{Berger:2006db}
\begin{eqnarray} 
\Gamma =3H(-1-w_\Lambda)
+2{{\dot{L}_\Lambda}\over {L_\Lambda}}\;.
\label{rates}
\end{eqnarray}
In addition to providing the connection between the various rates and the 
equation of state, this also establishes that the interaction should be 
of the size of $H$. 

Using the above formulism one can show that (assuming $w_m=0$)
\begin{eqnarray}
w_\Lambda^{\rm eff}&=&-1+{2\over 3H}{{\dot{L}_\Lambda}\over {L_\Lambda}}\;.\, 
\\
w_m^{\rm eff}&=&-{\Gamma \over {3H}}{{\Omega_\Lambda}\over 
{(1-\Omega_\Lambda)}}\;.
\label{general}
\end{eqnarray}
Then one obtains from Eq.~(\ref{diffgen}) the differential equation
\begin{eqnarray}
{{d\Omega_\Lambda}\over {dx}}=3\Omega_\Lambda(1-\Omega_\Lambda)
\left [1-{2\over 3H}{{\dot{L}_\Lambda}\over {L_\Lambda}}
-{\Gamma \over {3H}}{{\Omega_\Lambda}\over 
{(1-\Omega_\Lambda)}}\right ]\;.
\label{diffgeneral}
\end{eqnarray}
This evolution equation is general and is obviously most useful when an 
assumption for the holographic principle and for the interaction are specified.
For a constant equation of state together with a holographic principle 
the evolution equation becomes
\begin{eqnarray}
{{d\Omega_\Lambda}\over {dx}}=3\Omega_\Lambda
\left [1+w_\Lambda \Omega_\Lambda
-{2\over 3H}{{\dot{L}_\Lambda}\over {L_\Lambda}}
\right ]
\label{diffgenw}
\end{eqnarray}
On the other hand 
if a constant equation of state and the interaction is specified,
one can eliminate the explicit dependence on $\dot{L}_\Lambda/L_\Lambda$
and obtain
\begin{eqnarray}
{{d\Omega_\Lambda}\over {dx}}=3\Omega_\Lambda(1-\Omega_\Lambda)
\left [-w_\Lambda
-{\Gamma \over {3H}}{{1}\over 
{(1-\Omega_\Lambda)}}\right ]\;.
\label{diffgeni}
\end{eqnarray}

Finally the evolution of the Hubble parameter is given by
\begin{eqnarray}
{\dot{H}\over {H^2}}={1\over H}{dH\over dx}=
-{3\over 2}(1-\Omega_\Lambda)-{\Omega_\Lambda \over H}
{{\dot{L}_\Lambda}\over {L_\Lambda}}+{\Gamma \over {2H}}
\Omega_\Lambda\;.
\label{accelgen}
\end{eqnarray}
Equations (\ref{diffgeneral}) and (\ref{accelgen}) then determine a set of 
equations that give $\Omega_\Lambda(z)$ and $H(z)$ where $z$ is the redshift, 
and include the usual $\Lambda$CDM model as a special case.

\section{Interacting Dark Energy Model}

The $\Lambda$CDM model is defined in terms of the preceding formalism
as
\begin{eqnarray}
{{\dot{L}_\Lambda}\over {L_\Lambda}}=0\;, \qquad 
\Gamma=0\;. 
\end{eqnarray}
The equations of state of matter and dark energy are $0$ and $-1$ 
respectively, 
and the model predicts an evolution of the Hubble parameter as 
\begin{eqnarray}
H(z)=H_0\left [(1+z)^3\Omega_{m,0}+\Omega_{\Lambda,0}\right ]^{1/2}\;,
\label{LCDM}
\end{eqnarray}
where $\Omega_{m,0}$ and $\Omega _{\Lambda,0}$ are the present values of matter
and dark energy. A good fit to the supernova data is given by 
$\Omega_{m,0}\approx 0.3$ and $\Omega_{\Lambda,0}\approx 0.7$.

Various physical assumptions have appeard for the holographic condition on 
$\dot{L}_\Lambda/L_\Lambda$. The cosmological constant ($\Lambda$CDM) requires
$\dot{L}_\Lambda/L_\Lambda=0$. Other conditions involve invoking a holographic
arguement such as relating the length scale to various physical horizons. For 
example, one can relate $L_\Lambda$ to the Hubble horizon (HH), the particle
horizon (PH), or the future horizon (FH) for which
\begin{eqnarray}
R_{\rm HH}&=&1/H\;, \nonumber \\
R_{\rm PH}&=&a\int_0^t{dt\over a}=a\int_0^a {da\over Ha^2}\;, \nonumber \\
R_{\rm FH}&=&a\int_t^\infty {dt\over a}=a\int_a^\infty {da\over Ha^2}\;.
\end{eqnarray}
These yield 
\begin{eqnarray}
{1\over H}{{\dot{L}_\Lambda}\over {L_\Lambda}} 
&=&-{\dot{H}\over H^2}\quad {\rm (HH)}\;, \nonumber \\
&=&1+{\sqrt{\Omega_\Lambda}\over c}\quad {\rm (PH)}\;, \nonumber \\
&=&1-{\sqrt{\Omega_\Lambda}\over c}\quad {\rm (FH)}\;,
\label{holocase}
\end{eqnarray}
The HH condition implies
\begin{eqnarray}
{{d\Omega_\Lambda}\over {dx}}=0\;.
\end{eqnarray}
so that $\Omega_\Lambda$ (and $\Omega _m$) is constant.
This scenario has been shown to be
inconsistent with the observational data\cite{Hsu:2004ri} in the absence 
of any interaction. However as we now proceed to show, if a constant 
rate of decay $\Gamma$ of dark energy into matter is assumed, 
one can recover a good fit to the data.

The PH and FH conditions have been considered more recently, and have been 
exploited\cite{Li:2004rb,Kim:2005at} to attempt to 
achieve a fixed point solution for $\Omega_\Lambda$.
It has been 
shown that the FH condition
is needed if one is to obtain a value of $\Omega _\Lambda$ 
that is suitably close to the valued obtained from fits to the supernova data.
However assuming that $\Omega _\Lambda$ has acquired a value suitably close
to its fixed point value does not guarantee a good fit to the data. In fact,
by assuming PH or FH as the holographic condition and an interaction as a 
function on $\Omega _\Lambda$, the quantity $\dot{H}/H^2$ would be a function 
of $\Omega _\Lambda$ alone as in Eq.~(\ref{accelgen}). If also 
$\Omega _\Lambda$ is approximately constant because it is near its fixed 
point, then the universe is characterized by two components with 
constnat equations of state $w_m^{\rm eff}$ and $w_\Lambda ^{\rm eff}$. 

Consider the case where a holographic condition associates the dark energy 
length scale with the Hubble parameter (HH) and the interaction is set equal 
a constant of the same scale as the Hubble constant,
\begin{eqnarray}
{{\dot{L}_\Lambda}\over {L_\Lambda}}=-{{\dot{H}}\over H}\;, \qquad 
\Gamma=\kappa H_0\;. 
\label{modeldef}
\end{eqnarray}
Here we require $\kappa $ to be an order one constant.
As pointed out in Ref.~\cite{Hsu:2004ri} the first condition in 
Eq.~(\ref{modeldef}) in the absence of any interaction ($\Gamma =0$) 
gives rise to a 
description of the universe inconsistent with the observational data. In this
circumstance, the holographic condition implies that
the cosmological evolution will appear as matter-dominated ($w=0$) at all
times.  However
when the interaction (decay) of the dark energy in Eqn.~(\ref{modeldef})
is included, one can obtain
a suitable fit.
The Hubble parameter satisfies 
\begin{eqnarray}
{{H(z)}\over {H_0}}=
\left (1-{\kappa \over {3r_0}}\right )(1+z)^{3/2}+{\kappa \over {3r_0}}
\;,
\label{modelH}
\end{eqnarray}
where $r_0=\Omega_{m,0}/\Omega_{\Lambda,0}$.
This evolution of the Hubble parameter exhibits the needed features to 
agree with observations: for large $z$ a characteristic $(1+z)^{3/2}$ behavior
of a matter-dominated era with an approach at smaller $z$ 
to a constant $H(z)$ characteristic 
of a de Sitter phase. A good fit is obtained when
$\kappa/(3r_0) \approx 0.62$ (see Section IV). Comparing to the $\Lambda$CDM
solution in 
Eq.~(\ref{LCDM}), the best fits result for parameters {\it roughly} equal
to $\Omega _{\Lambda ,0}=2/3$ and $\Omega _{m,0}=1/3$ in the $\Lambda$CDM
and similarly $\kappa/(3r_0)=2/3$ in the interacting dark energy model
(the parameter $\kappa/(3r_0)$ in the 
interacting dark energy model plays an analogous role to that of
$\Omega_{\Lambda ,0}$ in the $\Lambda$CDM). Nevertheless the best fit value
for $\kappa/(3r_0)$ in the interacting dark energy model is not equal to the 
best fit value of $\Omega_{\Lambda,0}$ in the $\Lambda$CDM.

The characteristics of the model can be understood from 
\begin{eqnarray}
{\dot{H}\over {H^2}}={3\over 2}\left (-1+{\Gamma \over {3H}}
{{\Omega_\Lambda}\over {1-\Omega_\Lambda}}\right )
+{1\over {2(1-\Omega_\Lambda)}}{{d\Omega_\Lambda}\over {dx}}\;.
\label{accelgen2}
\end{eqnarray}
A constant $\Omega_\Lambda$ causes the last term to vanish and the 
$1/r=\Omega_\Lambda/(1-\Omega_\Lambda)$ factor
is constant and equal to its present value $1/r_0$. Then
\begin{eqnarray}
{\dot{H}\over {H^2}}={3\over 2}\left (-1+{\Gamma \over {3H}}
{{1}\over {r_0}}\right )\;,
\label{accelgen3}
\end{eqnarray}
Then $H(z)$ clearly exhibits a transition between a matter-dominated era
($\dot{H}/H^2=-3/2$) and a constant $H$ era 
($\dot{H}/H^2=0$ when $H_\infty = \kappa H_0/(3r_0)$ for $\Gamma =\kappa H_0$).
At early times the interaction is negligible, and the holographic condition 
enforces an expansion of the Universe that appears as matter-dominated. When 
the Hubble parameter becomes sufficiently small, the interaction becomes 
effective and a fixed point solution is reached. 

While the interacting dark energy
model predicts a constant $\Omega_\Lambda$ the coincidence problem
emerges as the choice of scale $\Gamma=\kappa H_0$. The quantity $\kappa$ 
is an order one to give a good fit to all observational data.

The holographic condition $\dot{L}_\Lambda/L_\Lambda=-\dot{H}/H$ implies the
effective equations of state are equal,
\begin{eqnarray}
w_m^{\rm eff}=w_\Lambda^{\rm eff}&=&-{\Gamma\over {3H}}
{\Omega_\Lambda \over {(1-\Omega_\Lambda)}}\;, \nonumber \\
&=&-{\Gamma\over {3Hr_0}}\;.
\end{eqnarray}
With the interaction $\Gamma =\kappa H_0$ one recognizes that the common
equation of state for the matter and dark energy components varies from $0$ to 
$-1$ as the Hubble parameter decreases from large values in the early Universe
to its asymptotic value ($H_\infty = \kappa H_0/(3r_0)$).

\section{Quantitative comparison of the model with observational data}

In this section we compare the interacting dark energy model predictions 
from Eq.~(\ref{modelH}) with supernova data. 
Comparison of the observational data with holographic models with 
{\it noninteracting} dark energy was 
performed in Refs.~\cite{Huang:2004wt} and \cite{Zhang:2005hs}. The holographic
condition based on the Hubble horizon (HH) is grossly inconsistent with the
data as shown by Hsu\cite{Hsu:2004ri}, but other holographic models based on 
the future event horizon (FH) can be made consistent with the data.

The luminosity distance is defined as 
\begin{eqnarray}
d_L(z)=(1+z)H_0^{-1}\int _0^z {{dz^\prime}\over {E(z^\prime)}}\;,
\end{eqnarray}
where $E(z)=H(z)/H_0$. In the interacting dark energy model, $E(z)$ is 
obtained from the integration of Eq.~(\ref{diffgen}). The observational data
for the supernovae is expressed in terms of an apparent magnitude and redshift.
Assuming the supernovae have the same absolute magnitude $M$, then
the extinction-corrected distance moduli is given by
\begin{eqnarray}
\mu(z)=5\log_{10}(d_L(z)/{\rm Mpc})+25\;.
\label{apparent}
\end{eqnarray}

Using the calculated distance moduli in Eq.~(\ref{apparent}) for an interacting
dark energy model and the supernovae data\cite{Riess:2004nr}, 
one can perform a $\chi$-squared 
fit,
\begin{eqnarray}
\chi^2=\sum _i {{\left [\mu_{\rm obs}(z_i)-\mu_{\rm mod}(z_i)\right ]^2}
\over \sigma_i^2}\;,
\label{chi2}
\end{eqnarray}
where the sum runs over the supernova data points and 
where the $\sigma_i$ are the experimental errors in each observation.
The $\Lambda$CDM model gives a $\chi^2$ of 178 for the 157 data points in the 
Gold data sample\cite{Riess:2004nr} for $\Omega_{m,0}=0.27$ and 
$\Omega_{\Lambda,0}=0.73$. 

A comparison of the supernova
data with the interacting dark energy model is shown in Fig.~1.
Taking the Hubble constant as $H_0^{-1}=2997.9h^{-1}$~Mpc,
the best fit values for $r_0/\kappa$ is $0.54$ and for the Hubble constant is
$h$ =0.64.
The $1\sigma$ and $2\sigma$ contours 
obtained when varying the parameters $h$ 
and $r_0/\kappa $ are shown in Fig.~2.

\begin{figure}[h]
\centerline{
\mbox{\includegraphics[width=3.50in]{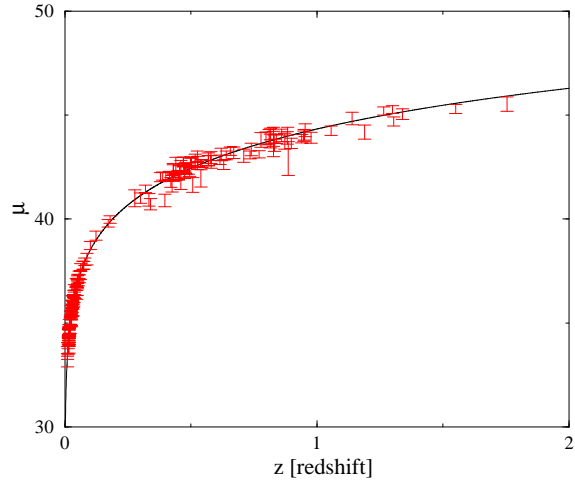}}
}
\caption{Comparison of the supernova data with the model prediction for 
$\mu(z)$ with $r_0/\kappa=1/2$. The data shown in the Gold data from 
Ref.\cite{Riess:2004nr}. 
The best fit occurs for $h$=0.64 for which
the $\chi^2$ is 179 for 157 degrees of freedom.}
\label{overView}
\end{figure}

The luminosity distance has dimensions of length and therefore scales with 
the inverse Hubble constant $H_0$. This means the model curve for $\mu(z)$ can
be shifted up or down by adjusting $H_0$.
The interacting dark energy model is also consistent with 
CMB measurement of the shift 
parameter\cite{Pearson:2002tr,Kuo:2002ua,Spergel:2003cb,Spergel:2006hy} and 
measurements of large scale structure by SDSS\cite{Eisenstein:2005su}.
Since the large $z$ behavior of the interacting dark energy model is the same
as the $\Lambda$CDM, the predicted value of shift parameter 
\begin{eqnarray}
{\cal R}=\sqrt{1-\Omega_\Lambda^0}\int _0^{z_{\rm dec}}
{{dz^\prime}\over {E(z^\prime)}}\;.
\label{shift}
\end{eqnarray} 
is easily accommmodated with a choice of $r_0/\kappa\approx 1/2$ and
$\Omega_{\Lambda,0}\approx 0.7$. 
The utility of 
the shift parameter ${\cal R}$ and the $A$-parameter measured by SDSS is that
they are independent 
of $H_0$\footnote{The shift parameter and the $A$-parameter have been used
previously to constrain models of holographic dark energy in 
Ref.~\cite{Zhang:2005hs}.}.
Consequently the 
gross geometric features of this model agree with the expansion history
predicted by the $\Lambda$CDM. An important question remains as to whether the
detailed observations of the large scale structure can be made consistent with
the matter density predicted at early times in 
this model (or some variant). 

\bigskip
\begin{figure}[h]
\centerline{
\mbox{\includegraphics[width=3.50in]{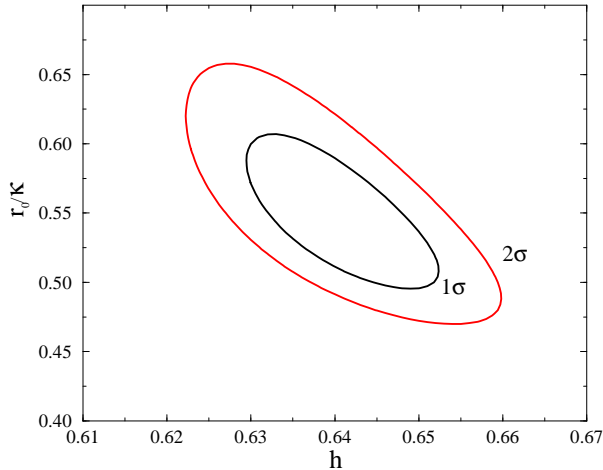}}
}
\caption{The $\Delta \chi ^2=2.30,6.17$ 
contours ($1\sigma,2\sigma$) for the parameters $h$ and $r_0/\kappa$ in
the interacting dark energy model.}
\label{fit}
\end{figure}
\bigskip

\section{Summary and Conclusions}

The $\Lambda$CDM model provides an explanation for all observational data. 
However there remain a number of important issues it must confront. Given our
lack of understanding of the dark energy, one can ask if there are other 
simple physical properties that dark energy might have that could equally well
account for the data. 

In this paper we have shown that a model of dark energy with a holographic 
condition relating the dark energy to the Hubble parameter and a constant 
interaction of size roughly equal to  
the Hubble constant $H_0$ can give a good fit to 
data. This model is characterized by a constant ratio of $\Omega _m$ to 
$\Omega_\Lambda$. Since the ratio of matter to dark energy is constant due to 
the holographic condition and the effective equations of state are equal, 
one can view the evolution of the model as one comprised of one component whose
effective equation of state varies between $0$ when the decay of the 
dark energy into matter is negligible to an asymptotic value of $-1$ when the 
interaction become important. The transition between one regime and the other
occurs when the interaction rate is comparable to the Hubble parameter. 

We have shown that the expansion history of the Universe can be reproduced
with a model of interacting dark matter. 
For the specific model presented here the 
functional dependence $H(z)$ differs slightly 
between the $\Lambda$CDM and the 
model discussed in this paper. Future data from SNAP may be useful in 
discriminating between them. Finally information that goes beyond the 
recorded expansion history of the Universe may be useful for ruling out this
kind of model. 

\section*{Acknowledgments}
This work was supported in part by the U.S.
Department of Energy under Grant No.~DE-FG02-91ER40661.

\end{document}